\documentclass[11pt]{JHEP3}

\usepackage{amssymb}
\usepackage{graphics}
\usepackage{epsfig}
\usepackage{graphicx}
\usepackage{subfigure}
\usepackage{bm}

\def\beq{\begin{equation}}
\def\eeq{\end{equation}}
\def\baq{\begin{eqnarray}}
\def\eaq{\end{eqnarray}}

\def\hnl{h_{\rm NL}}
\def\fnl{f_{\rm NL}}

\def\gnl{g_{\rm NL}}

\def\taunl{\tau_{\rm NL}}
\def\hnl{h_{\rm NL}}

\def\q{{\bf q}}
\def\x{{\bf x}}
\def\k{{\bf k}}
\def\d{{\rm d}}
\def\zg{\zeta_{\rm G}}
\def\ds{\delta\sigma}
\def\dss{\delta\sigma_{\rm s}}
\def\dsl{\delta\sigma_{\rm L}}

\def\bea{\begin{eqnarray}}
\def\eea{\end{eqnarray}}
\def\la{\langle}
\def\ra{\rangle}
\def\bkone{\mathbf k_1}
\def\bktwo{\mathbf k_2}

\def\anl{A_{\rm NL}}

\title{A non-Gaussian landscape}

\author{Sami Nurmi\,$^{a}$\footnote{sami.nurmi@helsinki.fi} , Christian T.~Byrnes\,$^{b}$\footnote{ctb22@sussex.ac.uk} , Gianmassimo Tasinato\,$^{c}$\footnote{gianmassimo.tasinato@port.ac.uk}\\
\\{\small $^{a}$\,
Department of Physics and Helsinki Institute of Physics, University
of Helsinki, P.O. Box 64, FIN-00014 University of Helsinki, Finland
\\
$^{b}$\,Astronomy Centre, University of Sussex, Brighton, BN1 9QH, UK\\
$^{c}$\,Institute of Cosmology \& Gravitation, University of
Portsmouth,  PO1 3FX, UK}}

\abstract { Primordial perturbations with wavelengths greater than
the observable universe shift the effective background fields in our
observable patch from their global averages over the inflating
space. This leads to a landscape picture where the properties of our
observable patch depend on its location and may significantly differ
from the expectation values predicted by the underlying fundamental
inflationary model. We show that if multiple fields are present
during inflation, this may happen even if our horizon exit would be
preceded by only a few e-foldings of inflation. Non-Gaussian statistics are especially affected: for example models of local non-Gaussianity predicting
$|\fnl^0|\gg 10$ over the entire inflating volume can have a
probability up to  a few tens of percent to generate a non-detectable
bispectrum in our observable patch $|\fnl^{\rm obs.}|\lesssim 10$.
In this work we establish systematic connections between the
observable local properties of primordial perturbations and the
global properties of the inflating space which reflect the
underlying high energy physics. We study in detail the implications
of both a detection and non-detection of primordial non-Gaussianity
by Planck, and discover novel ways of characterising the naturalness
of different observational configurations.}

\preprint{HIP-2013-02/TH}

\begin{document}
\section{Introduction}

Inflation is the leading scenario explaining the generation of
primordial perturbations. The imminent release of Planck data
 will provide crucial new information of the non-Gaussian
statistics of primordial perturbations. This will help to
efficiently discriminate between different realisations of inflation,
potentially ruling out whole classes of models. Future Large Scale Structure
surveys like Euclid will further improve our knowledge of the
statistics of primordial perturbations opening striking new insights
to the high energy physics behind inflation.

The observable primordial perturbations probe physics during the
last $N_{\rm obs.}\sim 60$ e-foldings of inflation after the horizon
exit of the largest observable modes. Inflation may however have
lasted much longer, so that the observable part of the universe
would constitute only a fraction of the entire inflating patch.
Long-wavelength perturbations generated before the horizon exit of
largest observable scales average to constants over our observable
patch, effectively shifting the local background field values. In
adiabatic single-field models it is well known that the
long-wavelength contributions amount to simply shifting the local
time coordinate \cite{Giddings:2010nc}. The situation is different
if more than one light dynamical scalar is present during inflation,
as in many models generating detectable non-Gaussianity. In this
case the evolution is non-adiabatic and the long-wavelength modes in
general shift both the time coordinate and the fields parameterizing
isocurvature directions. Shifts in the isocurvature directions
correspond to non-trivial changes of initial conditions which may
significantly affect the observational signatures and especially the
non-Gaussian statistics, irrespective of whether the isocurvature
perturbations persist until today.

The long-wavelength contributions are different in different parts
of the inflating region which generates variances in the physical
properties of patches smaller than the entire inflating region. This
leads to a landscape picture where the statistics of perturbations
within a horizon patch depend on the location of the patch
 itself \cite{Byrnes:2011ri}, as previously discussed in the curvaton scenario
   \cite{curvatonweb,Demozzi:2010aj}. The observable perturbations can of
course be described by an effective model covering only the last
$N_{\rm obs.}$ e-foldings of inflation in our horizon patch. The
background field values in the effective description are modulated
by a random long-wavelength contribution specific to our location in
the entire inflating patch. In the presence of isocurvature
directions the effective model can be non-trivially related to the
underlying fundamental inflationary model which covers the entire
inflationary epoch $N_{\rm tot.}$. Given a fundamental  model with
$N_{\rm tot.}> N_{\rm obs.}$ it is not possible to make firm
predictions for the observable signatures in our horizon patch but
one is led to consider probabilities of different signatures. It
turns out that especially in non-Gaussian models the differences
between the entire inflating patch, and of a random patch the size of
our observable patch,  can be significant. This is the case even if
the horizon exit of the largest observable modes would be preceded
by just a few e-foldings of inflation.

Since the fundamental model of inflation should be confronted with
theories of high-energy physics, it is of crucial importance to
develop a solid understanding of its relation to the effective
description of the last $N_{\rm obs.}$ e-foldings, which in turn can
be directly confronted with the observations made in our horizon
patch. In this work we will make systematic progress towards this
goal. We will establish novel relations between the observable
signatures and the form of the underlying inflationary physics. Our
model independent approach opens up interesting new possibilities to
address the naturalness of different non-Gaussian signatures. By
making use of these findings, we will establish new connections
between large and small non-Gaussianity and the structure of
inflationary physics.

\section{The curvature perturbation}

We concentrate on the class of generic single source models where
the dominant contribution to curvature perturbation is due to a
single scalar field, known as single-source inflation \cite{Suyama:2010uj}. This field can be an isocurvature degree of
freedom during inflation, e.g.~the curvaton scenario
\cite{curvaton,LUW} or modulated reheating \cite{mod1,mod2}. Such models can easily
generate observable non-Gaussianity. In the presence of isocurvature
directions, the long-wavelength fluctuations can not be removed by
merely shifting the time coordinate like in the adiabatic single
field case \cite{Giddings:2010nc}. Consequently, the long-wavelength modes can
have interesting observational consequences.

We assume the non-Gaussianity is generated on superhorizon scales
and neglect all contributions from subhorizon modes. While some
non-Gaussianity is in general also produced  by subhorizon physics,
these contributions are slow roll suppressed in models with
canonical kinetic terms. In this class of models the leading
contribution to observable non-Gaussianities is therefore
generated by superhorizon physics.

Neglecting the subleading subhorizon contributions, the curvature
perturbation $\zeta$ can be expressed in the local form
\beq
\label{ans}
\zeta(\x) = \zg(\x)+\frac{3}{5} \fnl^0
\left( \zg^2(\x)- \langle \zg^2\rangle \right)+\left(\frac{3}{5}\right)^2\gnl^0
\zg^{3}(\x)+\left(\frac{3}{5}\right)^3\hnl^0 \left( \zg^{4}(\x) -\langle \zg^4\rangle \right)
+\ldots\
\eeq
Here $\zg$ is a Gaussian field and the tree-level coefficients
$\fnl^0$, $\gnl^0$ etc. are constants. We are neglecting a possible
scale dependence of the non-linearity parameters, as discussed for
example in \cite{Byrnes:2010ft}.

The local Ansatz amounts to assuming that $\zeta$ is an analytic
function of a single Gaussian inhomogeneous field. To make this
manifest, we can equivalently rewrite (\ref{ans}) as
  \beq
  \label{taylor}
  \zeta(\x)
  =\sum_{n=1}^{\infty}\frac{N^{(n)}(\sigma_0)}{n!}\,\delta\sigma^n(\x)\
  ,
  \eeq
where $\sigma_{0}$ denotes a classical homogeneous solution of
equations of motion (and we understand that we remove from the even
powers in the previous formula the averages over the entire space,
to make the average of $\zeta$ over all space vanishing). This
corresponds to the familiar $\delta N$ expression \cite{starob85} if
$\delta\sigma(\x)\,=\,\zg(\x)/N'(\sigma_0)$ are chosen to represent
fluctuations on a spatially flat hypersurface soon after the horizon
exit. $N(\sigma)$ then measures the number of e-foldings from that
hypersurface to a uniform energy hypersurface at which $\zeta$ is
frozen to a constant value (which must exist provided that all
isocurvature perturbations decay).

To set the notation: the spectrum, bispectrum and trispectrum are
parameterized by
  \baq
  \langle\zeta_{\k_1}\zeta_{\k_2}\rangle &=&
  (2\pi)^3\delta(\sum\k_i)P_0(k_1)\ ,\\\nonumber
    \langle\zeta_{\k_1}\zeta_{\k_2}\zeta_{\k_3}\rangle &=&
  (2\pi)^3\delta(\sum\k_i)(\fnl^0 P_0(k_1)P_0(k_2)+{\rm perms.})\ ,\\\nonumber
  \langle\zeta_{\k_1}\zeta_{\k_2}\zeta_{\k_3}\zeta_{\k_4}\rangle &=&
  (2\pi)^3\delta(\sum\k_i)(\gnl^0P_0(k_1)P_0(k_2)P_0(k_3)+
  \\\nonumber& &+\,\taunl^0P_0(k_1)P_0(k_2)P_0(k_3)P_0(|\k_1+\k_2|)+{\rm perms.})\
  .
  \eaq
Since we are considering single source scenarios, the tree-level
parameters $\fnl^0$ and $\taunl^0$ are related by $\taunl^0 =(6/5)^2
(\fnl^0)^2$ \cite{Okamoto:2002ik,Byrnes:2006vq}.

\subsection{Long wavelength modes}

Let us consider the impact of long-wavelength fluctuations of the
field $\sigma$ in the expression for the curvature perturbation
(\ref{taylor}). We denote by $\sigma_{0}$ the classical homogeneous
solution which corresponds to the spatial average of $\sigma(\x)$
over the entire inflating patch. Fluctuations around the global
background can be divided into long and short wavelength components
with respect to the horizon scale of the observable universe $k_{\rm
obs.} = a_{\rm obs.} H_{\rm obs.}$
  \beq
  \delta\sigma(\x) = \int_{q > k_{\rm obs.}} \frac{\d\q}{(2\pi)^3}\,e^{i \q\cdot \x} \delta\sigma(\q)+\int_{q < k_{\rm obs.}} \frac{\d
  \q}{(2\pi)^3}\,e^{i\q\cdot\x}\delta\sigma(\q)\equiv \delta\sigma_{\rm s} (\x)+\delta\sigma_{\rm L} (\x)\ .
  \eeq
To first order in perturbations the long- and short-wavelength modes
are uncorrelated, $\langle\delta\sigma_{\rm
L}(\k)\dss(\k')\rangle=0$, and they both have vanishing ensemble
averages, $\langle\dsl\rangle=\langle\dss\rangle=0$. The ensemble
averages are assumed to coincide with the spatial averages computed
over the entire inflating patch, hence
$\langle\sigma(\x)\rangle=\sigma_0$.

Spatial averages computed over regions smaller than the inflating
patch in general differ from the ensemble averages. This is due to
the long-wavelength modes. For example, the average of $\dsl$ over a
spherical patch with radius $k_{\rm obs.}^{-1}$ and origin at $\x_0$
reads
  \baq
  \label{ds0}
  \langle\dsl(\x)\rangle_{{{\rm obs.}},\x_0}&\equiv&\frac{1}{V_{{\rm obs.}}}\int\d\x\,\theta(k_{\rm obs.}^{-1}-|\x-{\rm
  x}_0|)\,\dsl(\x)\\\nonumber
  &=&\frac{1}{V_{{\rm obs.}}}\int\frac{\d
  \q}{(2\pi)^3}\,\theta(k_{\rm obs.}-q)\ds(\q)\,e^{i\q\cdot{\rm
  x}_0}V_{{\rm obs.}}\left(1+{\cal
  O}\left(\frac{q^2}{k_{\rm obs.}^2}\right)\right)\\\nonumber
  &\simeq&\dsl({\bf x}_0)\ .
  \eaq
In order to pass from second to last line, we have assumed a scale invariant form for the mode functions,
$\ds(q)\propto q^{-3/2}$.

In a similar way one finds $\langle\dsl^n(\x)\rangle_{{\rm
obs.},\x_0}\simeq \dsl^n({\bf x}_0)$, so that the curvature
perturbation (\ref{ans}) in a patch the size of our observable
universe $k_{\rm obs.}^{-1}$ can be written as
  \beq
  \label{zeta0k}
  \zeta_{\rm obs.}(\k)\simeq\sum_{n=1}^{\infty}\frac{N^{(n)}(\sigma_0+\delta\sigma_{\rm L}(\x_0))}{n!}
  \,(\delta\sigma_{\rm s}(\k))^n\ .
  \eeq
Here $\dsl(\x_0)$ is a constant in $\langle...\rangle_{{\rm
obs.},\x_0}$ (but not in $\langle...\rangle$ since it depends on the
position $\x_0$). We have adopted a notation where $X_{\rm obs.}$
denotes a quantity $X$ evaluated in a patch of the size of our
observable universe.

The results depend on the location $\x_0$ of the patch as the
contribution of long-wavelength modes is different in different
regions of the inflating space. This position dependence limits our
ability to make precise predictions for observables in our horizon
patch \cite{Lyth:2007jh} if $N_{\rm tot.}>N_{\rm obs.}$. Instead of
assuming that we would occupy a ``typical'' location as in
\cite{Lyth:2007jh} we follow a top down approach and systematically
relate the probabilities for different observational configurations
in our patch to the form of the underlying fundamental inflationary
physics. Given an inflationary model with initial conditions set at
the beginning of inflation, we can make precise predictions only for
spatial averages over the entire inflating patch. These can
significantly differ from the averages in our horizon patch due to
the unknown long-wavelength contributions. However, given a model we
can unambiguously compute the probabilities for different
observational configurations in our horizon patch and thus work out
the possible observational imprints of the model. Going in the
opposite direction, we can also convert the observational
information into constraints on the fundamental high-energy physics
controlling the evolution of the entire inflating patch.

In this work we will systematically explore the consequences of the
long-wavelength modulations. We will establish novel connections
between the effective inflationary physics describing our observable
patch, and the fundamental inflationary physics which dictates the
statistics of perturbations over all of the inflating space and
directly reflects the underlying high-energy physics.

\section{Probability distributions of local observables}

If inflation lasted longer than $N_{\rm obs.}\sim 60$ e-foldings,
our observable universe constitutes an exponentially small fraction
of the entire inflating space. Patches of our horizon size $(a_{\rm
obs.} H_{\rm obs.})^{-1}$ located in different places of the
universe can then have different physical properties. This is a
consequence of the long-wavelength fluctuations which generate a
random modulation in the local expansion history. The modulation is
proportional to the number of e-foldings before the horizon exit of
our observable region
  \beq
  N_{\rm in.} \equiv {\rm ln}\,\frac{a_{\rm obs.}H_{\rm obs.}}{a_0 H_0}\ ,
  \eeq
where $t_0$ stands for the beginning of inflation. The longer the
duration of inflation, the more the unknown long-wavelength
contributions limit our ability to directly probe the underlying
high energy physics reflected in the statistics of inflationary
perturbations over the entire inflating space.

\subsection{Spectrum, bispectrum and trispectrum}

We assume that the fluctuations at horizon crossing $\delta\sigma =
\zeta_{\rm G}/N'$ in (\ref{taylor}) are Gaussian so that all
correlators can be expressed in terms of the two-point function
 \beq
  \label{ds2pt}
  \langle  \ds(\k) \ds(\k')\rangle = (2\pi)^3\delta(\k+\k')P_{\ds}(k)=(2\pi)^3\delta(\k+\k')\frac{2\pi^2}{k^3}{\cal
  P}_{\ds}\ .
  \eeq
For the canonical models of inflation we are concentrating on,
${\cal P}_{\ds}\simeq (H/2\pi)^2$ and the dependence of
$H(t,\sigma)$ on the field $\sigma$ is slow roll suppressed. In the
following we will neglect this dependence so that the two point
function of $\ds$ computed over an arbitrary subregion of the
inflating patch is given by the ensemble average (\ref{ds2pt}). In what follows we
will also ignore the slow roll suppressed scale dependence of ${\cal
P}_{\ds}$.

Using (\ref{zeta0k}), the spectrum of curvature perturbations
in a patch the size of our observable universe is given by
  \beq
  \label{Pexp}
  {\cal P}_{\rm obs.}
  ={\cal
  P}_0\left(\rule{0pt}{4ex}\right.1+\frac{12}{5}\fnl^0N'(\sigma_0)\delta\sigma_{\rm L}(\x_0)\left)\rule{0pt}{4ex}\right.\ ,
  \eeq
to first order in the long-wavelength field. ${\cal P}_0 =
N'{}^2(\sigma_0) {\cal P}_{\ds}$ and
$\fnl^0=5N''(\sigma_0)/6N'{}^2(\sigma_0)$ denote the ensemble
expectation values of the spectrum and the bispectrum amplitude,
which coincide with spatial averages over the entire inflating
patch.

Hence, the spectrum ${\cal P}_{\rm obs.}$ of fluctuations in a
horizon size patch depends on the location $\x_0$ of the patch
through the Gaussian field $\dsl(\x_0)$. This gives rise to a
landscape picture where the amplitude of curvature perturbations
measurable locally on patches of size $(a_{\rm obs.} H_{\rm
obs.})^{-1}$ fluctuates around the global average ${\cal P}_{0}$
according to a Gaussian probability distribution
  \beq
  \label{Pp}
  P({\cal P}_{\rm obs.})=(2\pi \sigma^2_{{\cal P}})^{-1/2}
  \,{\rm exp}\left(-\frac{({\cal P}_{\rm obs.}-{\cal P}_{0})^2}{2\sigma^2_{{\cal
  P}}}\right)\ .
  \eeq
Here the variance is given by
  \beq
  \label{sigmaP}
  \sigma_{{\cal P}}^2 = \left(\frac{12}{5}\fnl^0{\cal P}_{0}\right)^2
  N'{}^2(\sigma_0)\langle\dsl^2(\x_0)\rangle \simeq \left(\frac{12}{5}\fnl^0\right)^2{\cal
  P}^3_0 N_{\rm in.} \ .
  \eeq

In a similar way, using equation (\ref{zeta0k}) and working
to first order in $\dsl$, we can calculate the probability
distributions for $\fnl^{\rm obs.}$, $\taunl^{\rm obs.}$ and
$\gnl^{\rm obs.}$ measuring the non-Gaussian statistics in a horizon
size patch. The results are given by
  \baq
  \label{Pfnl}
  P(\fnl^{\rm obs.})&=&
  \frac{{\rm exp}\left(-\frac{(\fnl^{\rm obs.}-\fnl^{0})^2}{2\sigma^2_{\fnl}}\right)}{\sqrt{2\pi \sigma^2_{\fnl}}}
  \ ,\,\,\,\,\,
  \sigma_{\fnl}^2 = \left(\frac{9}{5}\gnl^0-\frac{12}{5}(\fnl^0)^2\right)^2{\cal
  P}_0 N_{\rm in.}\ ,\\
  \label{Pgnl}
  P(\gnl^{\rm obs.})&=&
  \frac{{\rm exp}\left(-\frac{(\gnl^{\rm obs.}-\gnl^{0})^2}{2\sigma^2_{\gnl}}\right)}{\sqrt{2\pi \sigma^2_{\gnl}}}
  \ ,\,\,\,\,\,
  \sigma_{{\gnl}}^2 = \left(\frac{12}{5}\hnl^0-\frac{18}{5}\gnl^0\fnl^0\right)^2{\cal
  P}_0 N_{\rm in.} \ ,
  \eaq
where $\hnl^0$ is defined as the fourth order coefficient in the
expansion of the curvature perturbation (\ref{ans}). The
non-linearity parameter $\taunl^{\rm obs.}$ in each patch is given
by $\taunl^{\rm obs.}= (6/5)^2(\fnl^{\rm obs})^2$ at scales
corresponding to the patch size. This is a result of the single
source form of the curvature perturbation $\zeta_{\rm obs.}
$(\ref{zeta0k}). At smaller scales the relation may become violated
by the loop corrections as discussed in \cite{Tasinato:2012js}.

Notice that the dependence of inflationary observables on the
location of the patch is closely associated to semiclassical loop
contributions to observable quantities, as we investigated in
\cite{Byrnes:2011ri,Tasinato:2012js} in collaboration with D. Wands.
The difference between the curvature perturbation $\zeta_{0}$ over
the entire inflating patch and the curvature perturbation
$\zeta_{\rm obs.}$ in our observable patch is due to the
long-wavelength modes $a_0 H_0 < k < a_{\rm obs.}N_{\rm obs.}$
generated before our horizon exit. Integrating over the unobservable
long wavelength modes to find $\zeta_{\rm obs.}$ (\ref{zeta0k})
amounts to computing radiative corrections to $\zeta_0$ at the scale
$k_{\rm obs.}=a_{\rm obs.}H_{\rm obs.}$. As explained in
\cite{Byrnes:2011ri,Tasinato:2012js}, the radiative corrections to
an $n$-point function of curvature perturbation consist of both the
usual loop corrections, where internal momenta are integrated over,
and of soft limits of higher order $n$-point functions where
external momenta become unobservably small. The soft contributions
generate variances for the $n$-point functions of $\zeta_{\rm obs.}$
causing the properties of a patch the size of our observable horizon
to depend on its location within the entire inflating space. It is
precisely the ramifications of this effect that we are investigating
in the current work.

As also explained in \cite{Byrnes:2011ri,Tasinato:2012js}, the next
to leading order long-wavelength corrections to ${\cal P}_{\rm
obs.}, \fnl^{\rm obs.}, \taunl^{\rm obs.}, \gnl^{\rm obs.}$
associated with higher powers of ${\cal P}_0$ are subdominant with
respect to the leading order corrections as long as the bare
non-linearity parameters $\fnl^0$, $\gnl^{0}$, $\hnl^{0}$ etc. are
not extremely large. In this work we are not considering such
extreme values and our first order analysis of the long-wavelength
corrections is therefore justified and self consistent.

\subsection{How large are the variances?}

The variances generated by the long-wavelength fluctuations are
uninterestingly small if there was basically no inflation before the
horizon exit of the largest observable modes, $N_{\rm in.} \lesssim
{\cal O}(1)$, and if the ensemble expectation values of the
non-Gaussian amplitudes are small, $|\fnl^0|\lesssim {\cal O}(1)$,
$|\gnl^0|\lesssim (\fnl^0)^2$. The more one deviates from either of
these conditions, the bigger the variances become.

The growing variances affect the comparison of inflationary models
and cosmological observations in two different ways. As the
variances of the theoretical predictions for the observables get
bigger than the observational sensitivity, the bounds on model
parameters get weakened, and projection of observational constraints
on  the space of model parameters may also become non-trivial.
Comparing the variance $\sigma_{\cal P}$ of the spectrum ${\cal
P}_{\rm obs.}$, given by equation (\ref{sigmaP}) to the 1-$\sigma$
error $\Delta{\cal P}/{\cal P} \simeq 0.1$ in $WMAP$ 7-years
parameter fits, we find that the theoretical uncertainty dominates
over the observational inaccuracy if
  \beq
  \sigma_{\cal P} > \Delta {\cal P}\qquad
  \Leftrightarrow\qquad \frac{|\fnl^0|}{50} \gtrsim \left(\frac{N_{\rm
  in.}}{300}\right)^{-1/2}\, .
  \eeq
Here we have set ${\cal P}_0=2.44 \times 10^{-9}$. For inflationary
modes featuring a few hundreds of e-foldings, $N_{\rm in.}\gtrsim
{\cal O}(10^{2})$, the variance becomes comparable to the
observational accuracy for $|\fnl^0|\gtrsim {\cal O}(10)$. This has
the effect of broadening the region of parameter space compatible
with observations. The growing variance increases the class of
models with different superhorizon scale properties, and underlying
inflationary physics, but degenerate predictions on observable
scales.

The variance of the spectrum can dominate over the expectation
value, $\sigma_{\cal P}\gtrsim {\cal P}_0$, if the inflationary
model features extreme non-Gaussianities, e.g. $|\fnl^0|\gg 10^2$,
and/or a very long period of inflation $N_{\rm in.}\gg 10^3$. In
both cases one should go beyond first order in the long-wavelength
perturbations $\dsl$. A more subtle issue is that the validity of
the entire semiclassical approach, justifying the use of the
expansion (\ref{taylor}), may become questionable in the regime
where fluctuations grow large: this is the regime investigated for
example in \cite{curvatonweb} in the case of the curvaton scenario.
In this work, we restrict ourselves to the regime of small
perturbations, tractable by a first order analysis in $\dsl$, where
the semiclassical approach should provide an accurate description of
the system. Consequently, we will not encounter situations where the
variance $\sigma_{\cal P}$ would dominate over the global background
${\cal P}_0$.

For the bispectrum the situation is different. Comparing the
variances of the spectrum (\ref{Pp}) and bispectrum (\ref{Pfnl}), we
obtain the relation
  \beq
  \label{rel_var}
  \frac{\sigma_{\fnl}^2}{(\fnl^0)^2} = \left(1-\frac{3}{4}\frac{\gnl^0}{(\fnl^0)^{2}}
  \right) \frac{\sigma_{{\cal P}}^2}{{{\cal
  P}_{0}^2}}\ .
  \eeq
Clearly, for models with  $|\gnl^0|\lesssim (\fnl^0)^2$ the relative
variances are of equal magnitude. Using the expected 1-$\sigma$
sensitivity of Planck, $\Delta\fnl = 5$ \cite{deltafnl}, we then
find that the theoretical uncertainty for $\fnl^{\rm obs.}$ becomes
comparable or dominates over the observational accuracy,
$\sigma_{\fnl}\gtrsim\Delta\fnl$, for
  \beq
  |\fnl^0| \gtrsim 5 \left(\frac{\sigma_{\cal P}}{{\cal P}_0}\right)^{-1}\
  .
  \eeq
As we will discuss below, this can happen in the regime
$\sigma_{\cal P} < \Delta {\cal P} $ where the theoretical variance
of the spectrum is still small compared to the observational
accuracy, and we may set ${\cal P}_{\rm obs.}\simeq{\cal P}_0$ to a
reasonable accuracy.

\begin{figure}[h!]
  \begin{center}
    \includegraphics[width=7 cm, height= 5 cm]{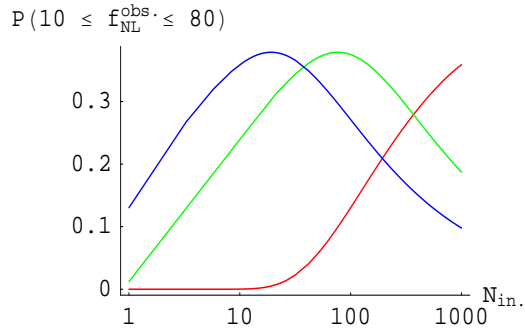}
  \end{center}
  \vspace{-0.5 cm}
  \caption{The probability to find $10 \leqslant \fnl^{\rm obs.} \leqslant 80$ in our observable part of the universe
  for different values of $\gnl^{0}$,
  assuming that the bispectrum averaged over the entire inflating patch
  vanishes $\fnl^0=0$. The probability is plotted against, $N_{\rm in.}$, the
  number
  of e-foldings from the beginning of inflation up to the horizon
  exit of our observable patch. The three curves
  correspond to $\gnl^0=10^4$ (red, rightmost),
  $\gnl^0=5\times 10^4$ (green, middle) and
  $\gnl^0=10^5$ (blue, leftmost), and we have set ${\cal P}_{0}=2.44 \times 10^{-9}$.}
  \label{fig:pfnl}
\end{figure}

On the other hand, for models with $|\gnl^0|\gg (\fnl^0)^2$, the
variance of the bispectrum can become very large even if our horizon
exit was preceded by just a few e-foldings of inflation. It may even
dominate over the background value, $\sigma_{\fnl}\gtrsim |\fnl^0|$,
effectively screening the underlying inflationary physics which
determines the global structure of the inflating space and reflects
the fundamental high energy physics behind inflation. While the
global average $\fnl^0$ represents the most probable value also in
patches the size of our observable universe, the probability for
observing a drastically different  bispectrum amplitude $\fnl^{\rm
obs.}$ becomes significant as the variance $\sigma_{\fnl}$ grows
large. This is illustrated in Fig.~\ref{fig:pfnl} for the specific
example where $\fnl^{0}=0$ globally but the model may still result a
detectable bispectrum $\fnl^{\rm obs.}$ in our observable patch.

The variance of the trispectrum may also be large. As we are
considering single source models, the observable $\taunl^{\rm obs.}$
calculated using (\ref{zeta0k}) is, at the scale $k_{\rm
obs.}=a_{\rm obs.}H_{\rm obs.}$ corresponding to the tree-level in
the observable universe, completely determined by $\fnl^{\rm obs.}$
through the relation $\taunl^{\rm obs.}=(6/5)^2(\fnl^{\rm obs.})^2$.
Its variance is then given by
  \beq
  \sigma^{2}_{\taunl} =
  4\left(\frac{6}{5}\right)^4(\fnl^0)^2\sigma_{\fnl}^2=4(\taunl^0)^2\frac{\sigma_{\fnl}^2}{(\fnl^0)^2}\
  .
  \eeq
It dominates over the background, $\sigma_{\taunl}^2\gtrsim
(\taunl^0)^2$, under the same conditions as $\sigma_{\fnl}^2\gtrsim
(\fnl^0)^2$. Looking at the equation (\ref{Pgnl}), we also find that
the variance of $\gnl$ may become large, $\sigma_{\gnl}^2\gtrsim
(\gnl^0)^2$, if the amplitude of the 5-pt function is large,
$|\hnl^0|\gg |\gnl^0|$. However, we will not consider this last
possibility in this work.

\section{Implications on inflationary physics}

The  cumulation of long-wavelength fluctuations can generate
significant variances in the small scale properties of primordial
perturbations across the inflating space. If inflation lasted much
longer than the $N_{\rm obs.}\sim 60$ e-foldings corresponding to
the horizon exit of the largest observable modes, this leads to a
landscape picture where the properties of our observable patch
depend on its location within the entire inflating space. This is
qualitatively similar to the special case of the curvaton web discussed
in \cite{curvatonweb}.

The superhorizon variation is not measurable and the observable
perturbations can   be described by an effective model of
inflation, covering only the last $N_{\rm obs.}$ e-foldings. The
effective model may however significantly differ from the
fundamental model for the entire inflating epoch. The Lagrangian of
the fundamental model is directly determined by the underlying
high-energy physics which may also provide a natural range of
initial conditions. In the effective model, the field values in the
Lagrangian are shifted by random long-wavelength fluctuations. This
may significantly affect the apparent form and predictions of the
theory in the presence of isocurvature directions.

The ensemble averages calculable from the fundamental model
correspond to the most likely predictions also in patches smaller
than the inflating volume, at least in the perturbative regime we
are considering here. The variances however make it possible to find
substantial differences between the local and global properties of
primordial perturbations. It is therefore of key interest to
carefully explore the relation between the observable local
quantities and the unobservable global inflationary perturbations
which reflect the underlying high energy physics. In particular, the
differences in non-Gaussian statistics over the entire inflating
patch and our observable universe may easily become large with
interesting consequences for observable quantities.

\subsection{Small primordial non-Gaussianity}

If Planck would not detect primordial non-Gaussianity, $|\fnl^{\rm
obs}|\lesssim 10$, then single-field slow-roll inflation would be
favoured as the minimal scenario consistent with observations.
However, nature might not have chosen the minimal setup. A small
bispectrum $|\fnl^{\rm obs}|\lesssim 10$ could also arise from
models with significant non-Gaussianities in the form of trispectrum
$|\gnl^{\rm obs.}|\gg (\fnl^{\rm obs.})^2$ (see \cite{Bernardeau:2003nw,Enqvist:2008gk} for examples of such
scenarios).  The signal would remain
unobserved by Planck if $|\gnl^{\rm obs.}|\lesssim 10^4$ but could
be revealed by future large scale surveys.

\begin{figure}[h!]
  \centering
    \includegraphics[width=15 cm, height= 5.5 cm]{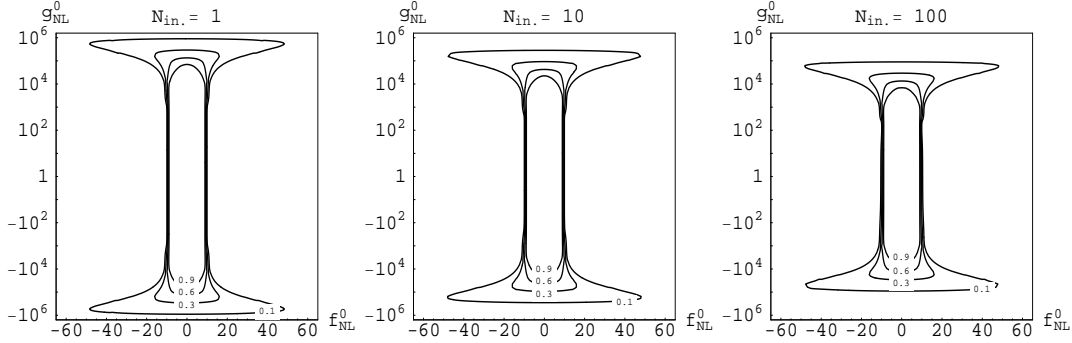}
  \caption{Probability to obtain a non-detectable bispectrum $|\fnl^{\rm obs.}|<10$ as a function of the
  ensemble expectation values
  $\fnl^0$ and $\gnl^0$ characterizing the underlying fundamental inflationary model. Three values of $N_{\rm in.}$ are shown corresponding to different total durations of the inflationary epoch.
   }
  \label{fig:fnlnondet}
\end{figure}

Using the landscape picture we introduced above, we can obtain novel
insights on the naturalness of such models. The variance
(\ref{Pfnl}) of the locally observable bispectrum amplitude
$\fnl^{\rm obs.}$ grows as $\sigma_{\fnl}\propto \gnl^0 N_{\rm
in.}^{1/2}$ assuming that $(\fnl^0)^2\ll \gnl^0$. Therefore, even if
the global bispectrum would be small $|\fnl^0|<10$, the probability
for obtaining an unobservably small bispectrum amplitude $|\fnl^{\rm
obs.}|\lesssim 10$ in our patch decreases as $\gnl^0$ grows or the
inflationary epoch becomes longer. This is illustrated in the
Fig.~\ref{fig:fnlnondet} which depicts the probability for
$|\fnl^{\rm obs.}|<10$ as a function of $\fnl^0$, $\gnl^0$ and
$N_{\rm in.}$, characterizing the fundamental inflationary model.
For the parameter range shown in the figure, the variance of the
spectrum is negligible compared to the observational accuracy
$\sigma_{\cal P}\ll \Delta {\cal P}$ and we have thus set ${\cal
P}_0\simeq {\cal P}_{\rm obs.}=2.44 \times 10^{-9}$. The variance of
trispectrum is also small so that $\gnl^{\rm obs.}\simeq \gnl^0$ to
a good precision.

The figure shows that models with $|\fnl^{0}|\lesssim 10$ and
$|\gnl^0|\lesssim 10^4$ have a high probability $P\gtrsim 0.9$ of
producing unobservably small bispectrum $\fnl^{\rm obs.}$ in our
horizon patch. The probability is not sensitive to the duration of
the inflationary epoch. We thus conclude (within the limits of
validity of our perturbative treatment) that a small bispectrum
$|\fnl^{\rm obs.}|<10$ is a natural outcome in inflationary models
with a Lagrangian leading to $|\fnl^{0}|\lesssim 10$ and
$|\gnl^0|\lesssim 10^4$.

On the other hand, Fig.~\ref{fig:fnlnondet} also shows that as the
trispectrum grows observable $|\gnl^0|\simeq |\gnl^{\rm obs.}|
\gtrsim 10^4$ the probability for an unobservable bispectrum
$|\fnl^{\rm obs.}|<10$ becomes smaller but not negligible. It is
also interesting to note that an unobservable bispectrum can be
generated even in models with a large global $|\fnl^0|\gtrsim 10$
even though the probability for this outcome is not higher than a
few tens per cent at most. In the regime of large trispectrum
$|\gnl^0|\gtrsim 10^4$ we also note that the probabilities become
sensitively dependent on the amount of inflation $N_{\rm in.}$
before our horizon exit. For example, the Fig.~\ref{fig:fnlnondet}
shows that the probability to find $|\fnl^{\rm obs.}|<10$ and
$|\gnl^{\rm obs.}|\gtrsim 10^5$ becomes very small for $N_{\rm
in.}\gg 10$ while it may be substantially larger for $N_{\rm
in.}\lesssim 10$. Therefore, to generate a signature $|\fnl^{\rm
obs.}|\lesssim 10, |\gnl^{\rm obs.}|\gtrsim 10^5$, inflation should
not have lasted much longer than the observable $N_{\rm obs.}$
e-foldings which represents a non-trivial tuning of model
parameters.

We emphasize that our approach is model independent, apart from our
restriction to the local form of non-Gaussianity (\ref{ans}). We
therefore cannot address the possible fine-tuning associated to
realizing specific values for $\fnl^0$ and $\gnl^0$ in a concrete
inflationary setup. In contrast, our approach reveals that,
independently of the details of inflationary physics, certain
configurations  $\fnl^0, \gnl^0$ in general fail to translate into
similar observational configurations $\fnl^{\rm obs.}, \gnl^{\rm
obs.}$ unless non-trivial constraints are placed on the duration of
inflationary epoch. We will address this model-independent tuning in
more detail below.

\subsection{Large primordial non-Gaussianity}

In the regime of detectable bispectrum by Planck, $|\fnl^{\rm
obs.}|\gtrsim 10$, the variance due to long wavelength modes can
have interesting effects. As the variance grows bigger than the
1-$\sigma$ accuracy of Planck, $\sigma_{\fnl}\gtrsim {\Delta\fnl} =
5$, the projection of observational constraints on to the model
parameters $\fnl^0$ and $\gnl^0$ becomes non-trivial.

\begin{figure}[h!]
  \centering
    \includegraphics[width=15 cm, height= 10 cm]{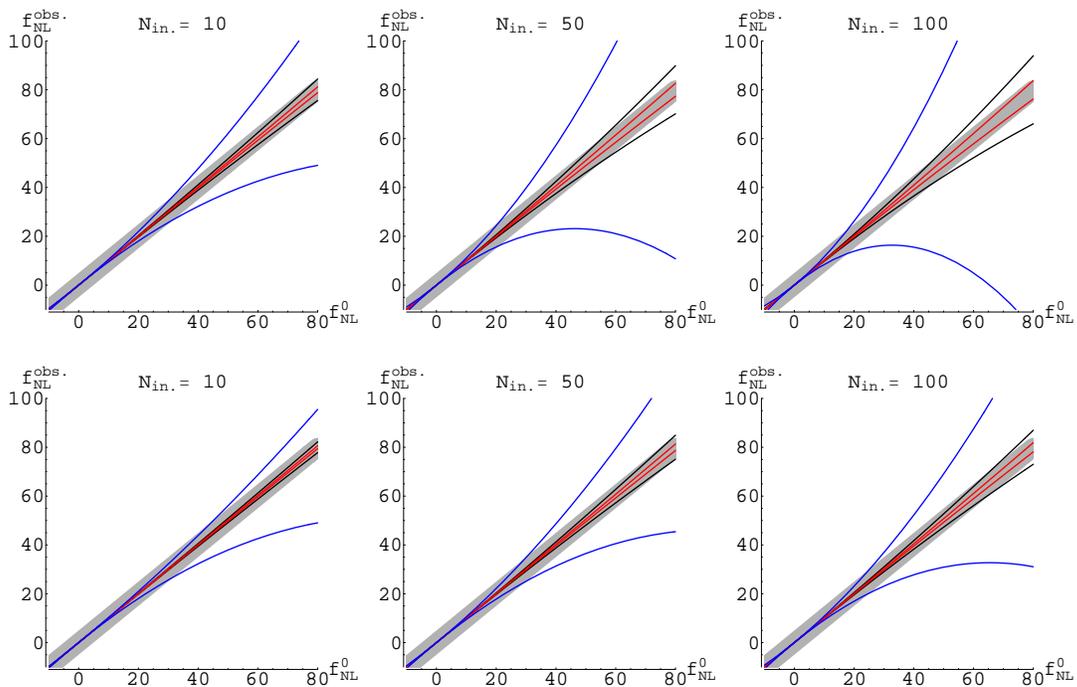}
  \caption{The observable bispectrum $\fnl^{\rm obs.}$ plotted against the theoretical
   expectation value $\fnl^0$ which reflects the underlying inflationary physics. The shaded area corresponds to
   the observational 1\,-\,$\sigma$ uncertainty of Planck $\Delta \fnl = 5$. The curves in the upper panel bound
   the theoretical 2\,-\,$\sigma$ regions $\fnl^{\rm obs.}=\fnl^{0}\pm 2\sigma_{\fnl}$ of the bispectrum amplitude
   $\fnl^{\rm obs.}$ in our observable patch. The lower panel shows the corresponding
   1\,-\,$\sigma$ regions $\fnl^{\rm obs.}=\fnl^{0}\pm \sigma_{\fnl}$. The three curves depict  $\gnl^0=0$ (black,
   middle), $\gnl^0=(\fnl^0)^2$ (red, innermost) and
   $\gnl^0=10 (\fnl^0)^2$ (blue, outermost). The results are shown for three choices of $N_{\rm in.}$, the number of e-foldings
   before the horizon exit of largest observable scales.
   The variance of the spectrum is smaller than the observational error $\sigma_{\cal P}/{\cal P}_{0} < 0.1$
   in the entire regime shown in the plots.}
   \label{fig:sigmafnl_deltafnl}
\end{figure}

This is illustrated in Fig.~\ref{fig:sigmafnl_deltafnl}. The figure
depicts the relation between the observable $\fnl^{\rm obs}$, and
the theoretical parameters $\fnl^0$ and $\gnl^0$ directly connected
to details of the underlying inflationary physics.
The variance $\sigma_{\fnl}$ grows along with the number of
e-foldings $N_{\rm in.}$ preceding the horizon exit of largest
observable modes. For $|\gnl^0|\gtrsim (\fnl^0)^2$, the
1\,-\,$\sigma$ region of $\fnl^{\rm obs.}$ can grow so large that a
detection of $\fnl^{\rm obs.}$ would not suffice to place tight
constraints on $\fnl^0$. Consequently, the observable value
$\fnl^{\rm obs.}$ may considerably deviate from the global average
$\fnl^0$. The same would also be true for the trispectrum $\gnl^{\rm
obs.}$ if $|\hnl^0|\gg |\gnl^0|$, i.e. the 5-pt function would be
large. Here we however do not consider this possibility further but
restrict ourselves to models where $|\hnl^0|\lesssim |\gnl^0|$, and
consequently $\gnl^{\rm obs.}\simeq \gnl^0$.

We recap that $\fnl^0$ corresponds to the tree-level amplitude of
the bispectrum computed for the inflationary model describing the
evolution of the entire inflating patch. The form of this
fundamental inflationary model is eventually dictated by the
underlying high-energy physics and we denote its Lagrangian
schematically by ${\cal L}_0$. On the other hand, the properties of
our observable patch can be parameterized by an effective model
${\cal L}_{\rm obs.}$ describing only the last $N_{\rm obs.}\sim 60$
e-foldings of inflation in our patch. Observables such as $\fnl^{\rm
obs.}$ then denote quantities calculable from the effective model
${\cal L}_{\rm obs.}$. In the semiclassical approach that we are
utilizing here, the field values of the effective Lagrangian are
shifted by the random long-wavelength contributions, as is clear
from equation (\ref{zeta0k}). From the point of view of the
effective model the shifts correspond to random changes of initial
conditions if isocurvature directions are present during inflation.
These can significantly affect the predictions, and especially the
non-Gaussian statistics, generating substantial differences in the
apparent form of the fundamental model ${\cal L}_0$ and the
effective description of our patch ${\cal L}_{\rm obs.}$. This has
interesting ramifications as only the fundamental model ${\cal
L}_{0}$ is the one directly reflecting the high energy physics
behind the inflationary epoch whereas the effective model ${\cal
L}_{\rm obs.}$ is the one directly constrained by observations.

As the Fig.~\ref{fig:sigmafnl_deltafnl} illustrates, the variance
$\sigma_{\fnl}$ can be large compared to the observational
1\,-\,$\sigma$ error $\Delta\fnl$ not only in models with
$|\gnl^0|\gg (\fnl^0)^2$ but also for $|\gnl^0|\ll (\fnl^0)^2$ if
$\fnl^0$ is large enough. As a curiosity, it is interesting to note
that the variance is accidentally suppressed around $\gnl^0\sim
(\fnl^0)^2$ due to cancellation\footnote{The vanishing of
$\sigma_{\fnl}$ for $\gnl^0=4/3 (\fnl^0)^2$ (the proportionality factor depends
 on the Ansatz for the curvature perturbation, in our case eq. (\ref{ans}))
 is an artefact of
neglecting the slow roll corrections; in general they would generate
a small $\sigma_{\fnl}$ even at $\gnl^0=4/3 (\fnl^0)^2$.} of the two
terms in (\ref{Pfnl}).
\begin{figure}[h!]
  \centering
    \includegraphics[width=15 cm, height= 6 cm]{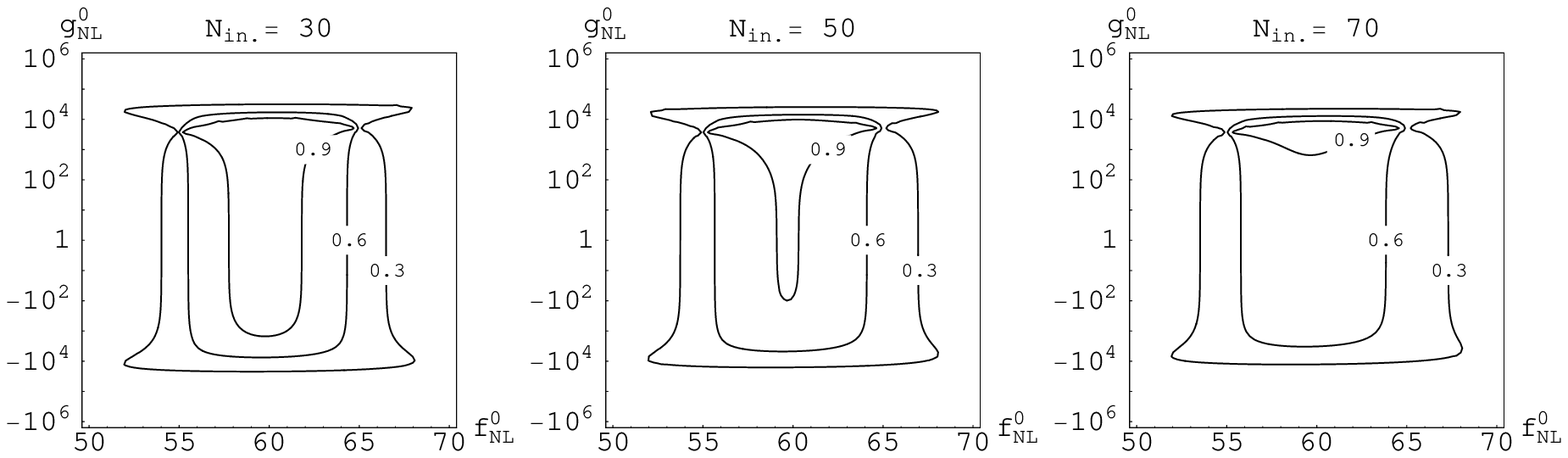}
  \caption{  The countours depict the probability to
  obtain a bispectrum in the range $55 \leq \fnl^{\rm obs.}\leq 65$ as a function of the model parameters $\fnl^0$ and $\gnl^0$.
  The results are shown for three choices of $N_{\rm in.}$, the number of e-foldings
   before the horizon exit of largest observable scales.
  The variance $\sigma_{\fnl}$ is accidentally suppressed for $\gnl^0\sim
  (\fnl^0)^2$. This enhances the probability to produce large $\fnl^{\rm obs.}$
  close to the ensemble average $\fnl^0$ of underlying inflationary physics.
}
  \label{fig:fnlgnlisland}
\end{figure}
In this regime the predictions of the model are not significantly
affected by the duration of inflation but the theoretical signature
$\gnl^0\sim (\fnl^0)^2$ is generically reflected with a high
probability by the observables $\gnl^{\rm obs.}$ and $\fnl^{\rm
obs.}$. This behaviour is illustrated in
Fig.~\ref{fig:fnlgnlisland}.

\begin{figure}[h!]
  \centering
    \includegraphics[width=15 cm, height= 5.5 cm]{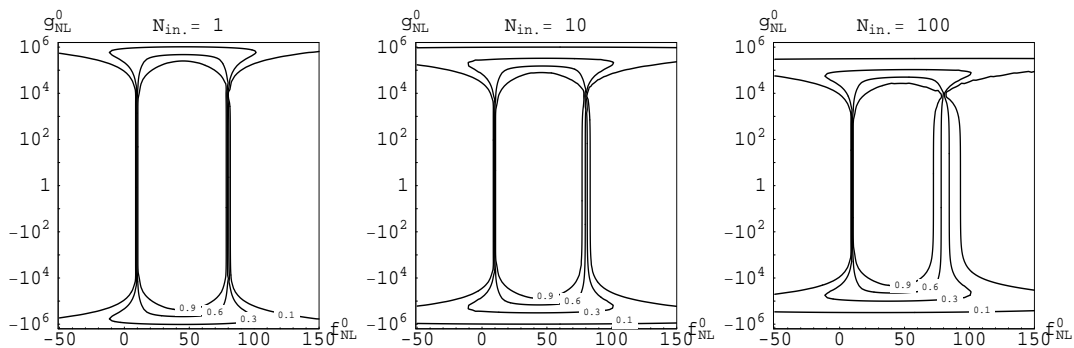}
  \caption{Probability to generate a bispectrum in the observable range $10 < |\fnl^{\rm obs.}| < 80$ as a function of the
  ensemble expectation values
  $\fnl^0$ and $\gnl^0$. Three values of $N_{\rm in.}$ are shown corresponding to different total durations of the inflationary epoch.
   }
  \label{fig:fnldet}
\end{figure}

When the variance $\sigma_{\fnl}$ grows comparable to $|\fnl^0|$ the
landscape effects become even more pronounced. In this regime the
observational signature becomes dominated by the long-wavelength
modes which may effectively screen the original signal $\fnl^0$,
reflecting the structure of the underlying inflationary physics. The
locally measurable $\fnl^{\rm obs.}$ then becomes strongly dependent
on the location within the inflating patch and $\fnl^{\rm obs.}$ may
significantly deviate from the ensemble expectation value $\fnl^0$.
This behaviour can be observed in Fig.~\ref{fig:fnldet} which
depicts the probability of generating an observable bispectrum as a
function of the model parameters $\fnl^0$ and $\gnl^0$.
Indeed, in the regime $|\gnl^0|\gg (\fnl^0)^2$ where
$\sigma_{\fnl}^2\gg (\fnl)^2$ the probability for obtaining
$10<\fnl^{\rm obs.}<80$ remains non-negligible even for $\fnl^0\gg
100$. The boundaries of the 60\% probability contours in the
$\fnl^0$\,-\, direction however do not significantly differ from the
bounds $10<\fnl^{\rm obs.}<80$ unless the inflationary period
becomes very long, $N_{\rm in}\gtrsim 10^2$. The slow tightening of
the upper bounds on $\fnl^0$ as a function of $N_{\rm in.}$, as
observed in the figure, is due to the growth of the variance
$\sigma_{\fnl}$ (\ref{Pfnl}). In the regime $\gnl^0\sim (\fnl^0)^2$
the variance is suppressed which explains the different behaviour
observed for these values in Fig.~\ref{fig:fnldet}.

For even longer periods of inflation $N_{\rm in.}\gg 10^2$ the
effects are expected to become even more significant. The analysis
of this regime is however beyond the scope of the current work as
our perturbative approach can not be directly applied to this regime
without carefully addressing convergence issues.

\subsection{Non-Gaussian signatures and fine-tuning}

If the variances grow large, the observational configurations become
sensitively dependent to $N_{\rm in.}$,  the amount of inflation
before our horizon exit. For the bispectrum this happens when
$|\gnl^0|\gg (\fnl^0)^2$ as can be clearly observed in
Figs.~\ref{fig:sigmafnl_deltafnl} and \ref{fig:fnldet}.

Indeed, our findings show that observable patterns $\{\fnl^{\rm
obs.},\gnl^{\rm obs.}\}$ with $|\gnl^{\rm obs.}|\gg (\fnl^{\rm
obs.})^2$ directly reflect the underlying theoretical signature
$\{\fnl^0,\gnl^0\}$ only if that signature was realized just before
the horizon exit of the observable scales. Consequently, as
$\{\fnl^0,\gnl^0\}$ characterise correlators over the entire
inflating space, this implies that inflation should not have lasted
much longer than the observable $N_{\rm obs.}$ e-foldings. If
inflation lasted much longer, $N_{\rm in.}\gg 1$, the rapidly
growing variance of the bispectrum $\sigma_{\fnl}$ effectively
screens the original signature $\{\fnl^0,\gnl^0\}$. The probability
of $\fnl^{\rm obs.}$ deviating from $\fnl^0$ then becomes
significant and the observational configuration $\{\fnl^{\rm
obs.},\gnl^{\rm obs.}\}$ no longer reflects the theoretical
signature $\{\fnl^0,\gnl^0\}$ dictated by the underlying
inflationary physics. Related considerations were also made in
\cite{Bernardeau:2003nw}.

\begin{figure}[h!]
  \centering
    \includegraphics[width=7 cm, height= 7 cm]{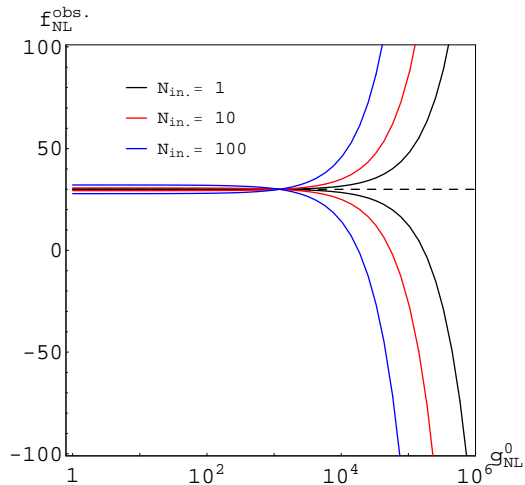}
  \caption{Configurations $|\gnl^0|\gg (\fnl^0)^2$ appear tuned as they become
  strongly modulated by long-wavelength modes if the horizon exit of the largest
  observable scales was preceded by even a few e-foldings of inflation.
  The dashed line depicts the value chosen for $\fnl^0$, the amplitude of the bispectrum over
  the entire inflating patch which reflects the underlying inflationary physics.}
  \label{fig:fgtuning}
\end{figure}

There is no a priori reason why the inflationary epoch should not
have lasted much longer than $N_{\rm obs.}$ e-foldings,
corresponding to the largest observable modes in the current
universe. Consequently, requiring $N_{\rm in.}$ to be small is a
non-trivial constraint indicating generic tuning associated to
configurations with large $\gnl^{\rm obs.}$. The range of e-foldings
$N_{\rm in.}$ after which an original  signature $|\gnl^0|\gg
(\fnl^0)^2$ gets screened by the long-wavelength modes is depicted
in Fig.~\ref{fig:fgtuning} as a function of $\gnl^0$.
Note that since we assume the amplitude of the 5-pt function is not
exceptionally large, $|\hnl^0|\lesssim (\fnl^0)^3$, the variance of
$\gnl^{\rm obs.}$ is negligible, $\sigma_{\gnl}^2\ll (\gnl^0)^2$,
for the range of parameters we are considering. Therefore to a good
precision we can approximate $\gnl^{\rm obs.}\simeq \gnl^0$.

Fig.~\ref{fig:fgtuning} shows that for signatures with $\gnl^{\rm
obs.}\sim 10^6$ and $(\fnl^{\rm obs.})^2\ll \gnl^{\rm obs.}$, the
bispecturm is dominated by the long-wavelength contributions already
if $N_{\rm in.}\gtrsim 1$. To translate a theoretical configuration
$\{\fnl^{0},\gnl^{0}\}$ with $|\gnl^0|\sim 10^6$ to the
corresponding observational signature $\{\fnl^{\rm obs.},\gnl^{\rm
obs.}\}$, the parameters $\fnl^0$ and $\gnl^0$ should thus be tuned
to the desired values less than one e-folding before the horizon
exit of the observable mode. If the values $\fnl^0$ and $\gnl^0$ are
set at an earlier stage, the observable $\fnl^{\rm obs.}$ has a high
probability to significantly differ from the theoretical expectation
$\fnl^0$. This suggests that such configurations are not very
natural -- unless the model under consideration naturally generates
$N_{\rm tot.}\sim N_{\rm obs.}$ e-foldings of inflation. For
$|\gnl^{\rm obs.}|\gtrsim 10^5$ the corresponding range of
e-foldings is given by $N_{\rm in.}\lesssim 10$ indicating a
somewhat milder tuning. For $|\gnl^{\rm obs.}|\gtrsim 10^4$ the
constraint is relaxed to $N_{\rm in.}\lesssim 100$.

We stress again that our discussion of the fine-tuning addresses
only the probabilities of finding observational configurations
$\{\fnl^{\rm obs.},\gnl^{\rm obs.}\}$ in our horizon patch given
that our horizon exit was preceded by some e-foldings of inflation.
The naturalness of realising certain values of $\fnl^0, \gnl^0$ in
specific inflationary models is another, model dependent issue which
may give rise to further tuning, or favour specific signatures. Our
general approach instead addresses the way how the theoretical
configurations $\fnl^0$ and $\gnl^0$ translate into measurable
properties in our observable patch. As we have shown, for the class
of models with $|\gnl^0|\gg (\fnl^0)^2$ these considerations suggest
that converting the value $\fnl^0$ into the corresponding
observational value $\fnl^{\rm obs.}\sim \fnl^0$ becomes unlikely
unless the horizon exit of our observable patch occurs soon after
the beginning of inflation. Imposing this requirement indicates a
generic source of non-trivial tuning in models with $|\gnl^0|\gg
(\fnl^0)^2$.

\section{Extension and applications of our results}

\subsection{Variance of the power spectrum in multi-source scenarios}

In this paper we have focused on single-source scenarios, and shown
under which circumstances there may be a significant variance of
observable quantities when making observations in a patch smaller
than the total inflated volume. Redoing this analysis in
multi-source scenarios goes beyond the scope of this work, but our
formalism makes it possible to do so, and here we give one small
example, thereby showing that our considerations remain interesting
in multi-source cases.

In a multi-source scenario, we must take into account the effect of
all relevant long wavelength fluctuations on to the short wavelength
modes. For the power spectrum, the lowest order correction comes
from the influence of the linear long wavelength modes on to the
linear short wavelength modes \bea \zeta_{\rm obs.}=
N^0_{A}\delta\sigma^A_{\rm s}+N^0_{AB}\delta\sigma^{A}_{\rm
s}\delta\sigma^{B}_{\rm L}+\cdots\,, \eea where summation convention
is used for the field indices, $A,B,\cdots$, over all field
components.

At tree level, and assuming that each component of the field
perturbations are uncorrelated, we may use \bea {\cal P}_0=N^0_A
N^{0,A}{\cal P}_{\sigma}\,, \qquad
\frac65\fnl^0=\frac{N_{AB}^0N^{0,A}N^{0,B}}{\left(N_C^0
N^{0,C}\right)^2}\,, \qquad \taunl^0=\frac{N^0_A N^{0,A}_B N^B_C
N^C}{\left(N^0_D N^{0,D}\right)^2}\,.  \eea Hence \bea \la
\zeta_{\rm obs.,\bkone} \zeta_{\rm obs.,\bktwo} \ra &=& N^0_A N^0_B
\la \delta\sigma_{{\rm s},\bkone}^A\delta\sigma_{{\rm
s},\bktwo}^B\ra + \left( N^0_A N^0_{BC} \la\delta\sigma_{{\rm
s},\bkone}^A\delta\sigma_{{\rm s},\bktwo}^B\ra \delta\sigma_{{\rm
L}}^C +1\;{\rm perm.} \right)  \\ {\cal P}_{\rm obs.}&=&{\cal P}_0
\left(1+2\frac{N^0_A
N^{0,A}_B\delta\sigma^{B}_{\rm L}}{N^0_C N^{0,C}}\right)\,, \\
\sigma^{2,\rm multi}_{\cal P}&=& 4\taunl^0 {\cal P}_0^3 N_{\rm
in.}\,. \eea Applying the Suyama-Yamaguchi (SY) inequality
$\taunl^0\geq (6\fnl^0/5)^2$ \cite{Suyama:2007bg}, it follows that
\bea \sigma^{2,\rm multi}_{\cal P}\geq
\left(\frac{12}{5}\fnl^0\right)^2{\cal P}_0^3 N_{\rm in.}=
\sigma^{2,\rm single}_{\cal P}\,. \eea
Hence for a given value of $\fnl^0$, the variance of the power
spectrum is larger than in single-source scenarios, showing that our
consideration about the difference between observables in the entire
inflated patch and our observable patch may become even more
important in multi-source scenarios.

\subsection{What will we observe if the single-source equality is broken by loops?}

Since we have assumed in most of this paper that we are in a
single-source scenario, it is interesting to consider how we can
observationally check whether this is the case. The well known test
is the equality $\taunl^{\rm obs.}=(6\fnl^{\rm obs.}/5)^2$ which
however is only valid at tree-level
\cite{Byrnes:2011ri,Sugiyama:2012tr}. In what follows we are mainly
considering quantities in our observable patch and for brevity we
omit the labels ${}^{\rm obs.}$ where no risk for confusion arises
and simply denote $\fnl\equiv\fnl^{\rm obs.}$ etc. Loop effects can
most easily break the equality when $\gnl$ is large, a scenario we
have extensively studied in this paper, making this particularly
relevant to study here. This would cause a deviation from the
equality when considering loop corrected quantities at scales
smaller than our horizon patch. However there is a generalised
equality between radiatively corrected quantities which include both
loop corrections where internal momenta are integrated over and soft
corrections where external momenta become small. This new equality
is valid to all orders in radiative corrections and it provides a
new test of inflation, as shown in \cite{Tasinato:2012js}.

A priori, in order to see a breaking of the equality, $\taunl$ is
the only parameter which we need to observe, regardless of whether
it is broken by loops in a single-source scenario, or by
multi-source effects. Here we will show that in practice if it is
broken by loops, we are essentially guaranteed to see all three
non-linearity parameters. The calculation is restricted to the one
loop level.

To be concrete, we assume that Planck can detect non-linearity
parameters with $2-\sigma$ confidence if $\fnl>10, \taunl>10^3,
\gnl>10^4$ \cite{Kogo:2006kh,Smidt:2010ra}.

Assuming that the SY inequality is always valid
\cite{Smith:2011if,Sugiyama:2012tr,Assassi:2012zq,Kehagias:2012pd},
to see a breaking of the equality with Planck we have to observe
$10^3\lesssim\taunl\lesssim10^4$ where the upper bound comes from
the current observational constraint. In this case
$30\lesssim|\fnl|\lesssim10^2$ if the equality holds and we would
clearly observe $\fnl$ as well. The key observation is that in order
for $\fnl$ not to be seen, $|\fnl|\lesssim 10$, we need
$\anl\equiv\taunl/(6\fnl/5)^2\gtrsim 10^3/10^2\sim10$, so the
equality needs to be strongly broken in this case. From
\cite{Tasinato:2012js} we find at one-loop level
\begin{equation}
\anl\equiv\left(\frac{5}{6}\right)^2\left(\frac{\taunl}{\fnl^2}\right)_{{\rm
tree}+{\rm loop}}\simeq1+4\left(\frac{\gnl}{\fnl}\right)_{\rm
tree}^2{\cal P}\ ,
\end{equation}
where the second term on the right hand side of the last equality is
responsible for breaking the equality. In order to get $\anl\gtrsim
10$ with $|\fnl|\lesssim 10$ we should then have $|\gnl|\gtrsim
3\times 10^5\gg\fnl^2$ at tree-level and hence the ``normal''
hierarchy between $\gnl$ and $\fnl$ must also be strongly broken. A
breaking of the equality must therefore come through an observably
large $\gnl$, which is close to the current upper limit. Such a
large tree-level $\gnl$ does not create problems with a loop
correction to either $\fnl$ or the power spectrum, and the
perturbative expansion in loops remains valid
\cite{Tasinato:2012js}.

Finally given that we require such a large value of $\gnl^{\rm
obs.}$, what is the probability that $\fnl^{\rm obs.}$ is too small
to be seen in our Hubble volume value? The answer can be seen in
Fig.~\ref{fig:fnlnondet}, for $|\gnl^{\rm obs.}|\simeq
|\gnl^0|\gtrsim 10^5$ the probability of obtaining $|\fnl^{\rm
obs.}|<10$ is suppressed below about $50$\% already for $N_{\rm
in.}\sim 1$ and it rapidly decreases as $N_{\rm in.}$ grows. We may
thus conclude that a breaking of the SY equality in single-source
inflation is likely to be accompanied by a measurement of all three
non-linearity parameters. It should then also be possible to observe
our equality $\taunl^{\rm rad.} = (6/5)^2(\fnl^{\rm rad.})^2$
between the radiatively corrected quantities $\taunl^{\rm rad.}$ and
$\fnl^{\rm rad.}$ which contain not only the loop corrections but
also the corrections from soft modes
\cite{Byrnes:2011ri,Tasinato:2012js}.

\section{Conclusions}

Primordial perturbations with wavelengths greater than the
observable universe contribute to background quantities in our
observable patch. The long-wavelength contributions vary across the
inflating patch and amount to random shifts of the local field
expectation values. This leads to a landscape picture where the
properties of our observable patch depend on its location. The
observable primordial perturbations constrain the form of the
effective model for the last $N_{\rm obs.}$ e-foldings of inflation
in our observable patch. Being affected by random long-wavelength
contributions the effective model may considerably differ from the
fundamental model of inflation which describes the entire inflating
patch and whose form is dictated by the underlying high energy
physics.

In this work we have shown that if multiple fields are present
during inflation, the long-wavelength contributions can be
significant even if our horizon exit would be preceded by only a few
e-foldings of inflation. Long-wavelength modes generate shifts along
isocurvature directions which correspond to non-trivial changes of
initial conditions in the effective model and can greatly affect the
resulting perturbations, especially the non-Gaussian statistics. In
particular, the impact of the long-wavelength modes can not be
deduced from the effective model itself but they reflect the form of
the underlying fundamental model of inflation. This is in sharp
contrast to pure single field slow roll models where the
long-wavelength effects are slow roll suppressed and amount to local
time shifts expressible in terms of spectral indices. As the
simplest example of multiple field models, we have in this work
concentrated on long-wavelength effects in the generic class of
single source models with the local type of non-Gaussianity. Many
theoretically interesting models, for example the curvaton and
modulated reheating scenarios in the limit of negligible inflaton
perturbations, fall into this class of models where perturbations
are generated by a single inhomogeneous field, despite the presence
of several light fields during inflation.

We have found novel general relations between the primordial
perturbations in our observable part of the universe and the
inflationary physics controlling the entire inflating patch. We have
investigated the consequences of both a detection and non-detection
of the primordial bispectrum $\fnl^{\rm obs.}$ by Planck. Our
results demonstrate that an eventual non-detection would not
strictly rule out models with a large bispectrum. Indeed if
inflation lasted longer than the observable $N_{\rm obs.}$
e-foldings, models with a bispectrum up to $\fnl^0\sim 40$ when
averaged over the entire inflating patch, can have a probability up
to few tens of per cents for yielding $|\fnl^{\rm obs.}|\lesssim 10$
in our observable patch. However, this is not a very generic feature
but requires a large trispectrum amplitude $|\gnl^{0}|\gtrsim 10^4$
and a value of $\gnl^0$ falling to a specific narrow range.

In case $\fnl$ is detected, the long-wavelength effects quite
generically have interesting implications for the interpretation of
the observation. If the inflationary model predicts $\gnl^0\gtrsim
(\fnl^0)^2$ across the entire inflating patch, we find that the
observable bispectrum $\fnl^{\rm obs.}$ has a high probability to
deviate from the global mean $\fnl^0$ even if there were only a few
tens of e-foldings of inflation before our horizon exit. As we have
shown in this work, the observational constraints on $\fnl^{\rm
obs.}$ therefore translate in a highly non-trivial way on the
parameters $\fnl^0,\,\gnl^0$, characterizing the underlying
inflationary physics.

We have also shown that inflationary models predicting a hierarchy
$|\gnl^0|\gg (\fnl^0)^2$ imply an inherent tuning of the total
duration of inflation irrespective of details of the inflationary
physics. This is because the variance of the bispectrum easily
becomes very large after a short period of inflation making it
unlikely to find an observable value $\fnl^{\rm obs.}$ which would
reflect the global mean $\fnl^0$. For $|\gnl^0|\sim 10^6$ this
happens if our horizon exit was preceded by just one e-folding of
inflation and for $|\gnl^0|\sim 10^5$ after about ten e-foldings. To
translate a theoretical configuration $|\gnl^0|\gg (\fnl^0)^2$ to
the corresponding observational signature, the total duration of
inflationary epoch should not be much longer than the observable
e-foldings $N_{\rm obs.}$. This tuning represents a highly
non-trivial constraint on inflationary physics.

It would be interesting to extend our results to generic multiple
field models where the long-wavelength effects could be even more
pronounced, as we have briefly discussed above. Another interesting
line of future work would be to consider non-Gaussianities generated
by subhorizon physics which generate non-trivial deviations from the
local form. Developing a solid understanding of the long-wavelength
effects would indeed be of key importance as it appears to play an
essential role in translating the new observational data into a
deeper understanding of fundamental high-energy physics.

\bigskip

\noindent {\bf Note added} Whilst this paper was being completed, a
related work appeared on the arxiv by Nelson and Shandera
\cite{Nelson:2012sb}.
 This work is conceptually similar to ours, since it also
studies the relation between our observable correlators to those in
a much larger patch in single-source scenarios and focusing on the
local model. However their focus is on the structure and hierarchy
of the non-Gaussian moments, while we focused on more concrete
calculations, such as explicitly finding how much $\fnl$ varies
under the influence of $\gnl$.  This paper and \cite{Nelson:2012sb}
are therefore complementary.

\acknowledgments{We thank
Jussi V\"{a}liviita and Matti Savelainen
for helpful discussions on statistics.
S.N. is supported by the Academy of Finland grant
257532. C.B. is supported by a Royal Society University Research
Fellowship. G.T. is supported by an STFC Advanced Fellowship
ST/H005498/1.  C.B. and G.T.  would like to
thank the theory group of Nordita, Stockholm  for their warm
hospitality.}

\end{document}